\documentclass[%
preprint,
superscriptaddress,
 amsmath,amssymb,
]{revtex4-2}

\usepackage{graphicx}
\usepackage{dcolumn}
\usepackage{bm}

\usepackage{xcolor}
\usepackage{booktabs} 
\usepackage{siunitx}
\usepackage{makecell}

\begin{document}

\preprint{}

\title{Impact of control signal phase noise on qubit fidelity}

\author{Agata Barsotti}
\email[Email: ]{agata.barsotti@phd.unipi.it} 
\affiliation{Dipartimento di Ingegneria dell'Informazione, Università di Pisa, Via Caruso 16, I-56122 Pisa, Italy}

\author{Paolo Marconcini}
\affiliation{Dipartimento di Ingegneria dell'Informazione, Università di Pisa, Via Caruso 16, I-56122 Pisa, Italy}
\affiliation{INFN Sezione di Pisa, Largo Bruno Pontecorvo 3, 56127 Pisa, Italy}

\author{Gregorio Procissi}
\affiliation{Dipartimento di Ingegneria dell'Informazione, Università di Pisa, Via Caruso 16, I-56122 Pisa, Italy}

\author{Massimo Macucci}
\affiliation{Dipartimento di Ingegneria dell'Informazione, Università di Pisa, Via Caruso 16, I-56122 Pisa, Italy}
\affiliation{INFN Sezione di Pisa, Largo Bruno Pontecorvo 3, 56127 Pisa, Italy}

\date{\today}

\begin{abstract}
{As qubit decoherence times are increased and readout technologies are improved, nonidealities in the drive signals, such as phase noise, are going to represent a crucial limitation to the fidelity achievable at the end of complex control pulse sequences. While prior studies have addressed the impact of reference oscillator phase noise on qubit performance, its interaction with realistic control pulses and its role in fidelity degradation have not been examined in sufficient detail. Furthermore, previous analyses grounded in the filter-function formalism identify high-frequency spectral components as the dominant source of fidelity loss, a conclusion that has also been used as a compelling marketing feature by some hardware vendors. Here we reevaluate this assertion by means of direct time-domain numerical simulations, in which phase noise realizations with a given power spectral density are applied to the carrier of realistic control pulse sequences, and the resulting qubit evolution is computed with Qiskit-Dynamics and averaged over multiple noise realizations. We show that the claimed dominance of high-frequency components originates from an oversight in interpreting the conversion between the phase-noise and the frequency-noise power spectral densities: at equal power, the components close to the Rabi frequency are by far the most detrimental, while those far from the carrier act only marginally, mainly through residual amplitude modulation. We performed an analysis using both two- and three-level models, the latter capturing leakage outside the computational subspace. We further show that, for realistic phase noise spectra of local oscillators, whose power is concentrated in the low-offset region, the fidelity loss is dominated by such high-power low-frequency components.
}

\end{abstract}

\maketitle

\section*{Introduction}
{The development of large-scale, reliable quantum computers is at present one of the main objectives of the scientific community and requires qubits with long relaxation ($T_1$) and coherence ($T_2^{\ast}$) times \cite{Nielsen_Chuang_2010,PhysRevApplied.3.044009,Burnett2019DecoherenceBO,Carroll:2021ltj}, long enough, in particular, for quantum error correction algorithms to be executed \cite{shor1995scheme,ryan2021realization,caune2024demonstratingrealtimelowlatencyquantum}.}

{Sources of decoherence are random physical processes in the environment surrounding the quantum processor, but also noise and fluctuations in the signals used to control qubits. Therefore, specifically understanding and reducing the effects of noise affecting control signals on decoherence are fundamental challenges within the overall development of quantum computers. Here, we focus, {{in particular,}} on the investigation of the action 
of phase noise, which is the prevalent nonideality (other forms of noise, such as thermal and shot noise, have in general a minor effect in this specific application). We will mainly deal with superconducting qubits, which represent one of the most promising implementations for quantum computers, due to their relatively long coherence times, as well as to the ease of manipulation \cite{10.1063/1.5089550,Huang:2020sce,GU20171,doi:10.1126/sciadv.abi6690}.}

{The impact of qubit control noise has traditionally been analyzed, within Quantum Control Theory \cite{10.1063/1.525634,Wiseman_Milburn_2009,Dong_2010, aroch}, by means of spectral overlap techniques, in which the effect of the noise is expressed as the overlap, in the Fourier domain, between its power spectral density and a spectral function determined by the applied control \cite{Kofman_2001,PhysRevLett.93.130406}; this idea underlies, in particular, the theory of Dynamical Error Suppression, which exploits sequences of appropriately designed pulses to mitigate the effects of dephasing \cite{PhysRevA.58.2733, Viola_1999,Uhrig_2007,Khodjasteh_2009,Biercuk_2009}, and was later formalized in terms of filter functions associated with the control pulse sequence \cite{filterfunction, Biercuk_2011,Green_2012,Green_2013, maloney}. Ball \textit{et al.}~\cite{ball} extended the filter-function formalism to the phase fluctuations of the reference oscillator used for pulse generation, by shifting such fluctuations onto the qubit: since the choice of the reference frame is arbitrary, phase fluctuations of the local oscillator can be treated as opposite phase oscillations of the qubit. Their approach was later applied to the phase and wide-band additive noise of a realistic oscillator~\cite{vandijk} and to the phase noise of cryo-CMOS oscillators~\cite{matsuoka}. {{Within}} their analysis, Ball \textit{et al.} asserted that the high-frequency components of the oscillator phase noise spectrum, although lower in amplitude, dominate qubit infidelity, a conclusion that has also {{resurfaced as a selling point by some hardware vendors.}}}

{The frequency dependence of the effect of control-field phase noise on gate fidelity has also been investigated for other qubit platforms. In particular, Jiang \textit{et al.}~\cite{jiang} studied the sensitivity of gate fidelity to laser phase and intensity noise in neutral-atom qubits, showing that fidelity is mainly degraded by spectral noise features located near the Rabi frequency, and that it can be improved by a proper choice of the Rabi frequency with respect to such features; related observations have been reported for trapped-ion gates subject to fast laser noise~\cite{nakav}. }


{One of the main contributions of the present work is a reevaluation of the statement by Ball \textit{et al.} about the dominance on fidelity loss of high-frequency phase noise components. Indeed, such result is a consequence of an oversight in the interpretation of the conversion between the phase-noise and the frequency-noise power spectral densities and of the subsequent calculations. Our direct time-domain numerical simulations confirm that high frequency components do not have a strong effect on fidelity, due to their limited amplitudes and the frequency selective nature of the qubit. At equal power, the most detrimental contribution to fidelity loss comes instead from the noise components close to the Rabi frequency, with the far-from-carrier components acting only marginally, mainly through residual amplitude modulation (AM)y. However, for the phase noise spectra of realistic local oscillators, in which the power is concentrated in the low-offset region, the overall fidelity loss turns out to be dominated by such high-power components, as well as components near the Rabi frequency. With respect to the approach by Ball \textit{et al.}, we take the rotating frame of the qubit as the reference and directly evaluate the interaction between the qubit and the noisy control pulses, also including the accumulation of the phase error of the reference clock with respect to the rotating frame. If correctly applied, the two approaches lead to substantially equivalent results, but ours makes the analysis of the contribution of each spectral component to the fidelity of specific final states more direct and intuitive, and it naturally accommodates more realistic qubit models and noise types beyond phase noise. In particular, it can be extended also to a three-level model, accounting for leakage outside the computational subspace.}

{Our approach is based on a two-step method. First, a colored phase noise process is synthesized from white Gaussian noise by linear spectral shaping (for example, see~\cite{Kasdin1995} for an application to
power-law noise); the specific finite-impulse-response realization adopted here is described in the next section.
Second, the resulting phase sequence is used as an input for a numerical simulation of the time evolution of the qubit state, performed with Qiskit-Dynamics~\cite{qiskit}
}
{(alternative tools exist in the literature, such as QuTiP~\cite{qutip}, used for example by Tong \textit{et al.}~\cite{tong} to study the effect of constant phase and frequency shifts and of other errors in the control signal). The final state reached by the qubit is compared with that from an ideal, noiseless evolution; simulations are repeated for 20 different realizations of the same phase noise process (changing the seed used to generate the pseudorandom sequence) and averaged, to obtain an estimate of the fidelity achievable for each considered noise PSD.}

\section*{Method}\label{sec_four}


{
In this section, we describe in more detail the numerical procedure used to generate the baseband phase noise sequences. 

The starting point is the specification of a desired frequency-domain behavior, which defines how the generated noise process should be shaped across frequency. The corresponding ideal impulse response is obtained by transforming this frequency response into the time domain. However, the resulting ideal impulse response is generally not finite in duration. Instead, it extends infinitely in both directions (i.e., it is noncausal and of infinite length).}
Such a filter cannot be implemented in practice, since only a finite number of coefficients can be stored and computed.

In practice, given a desired (real) amplitude function $A_d(\omega)$, a convenient {finite-dimensional} approximation of the desired filter is obtained through the Impulse Response Truncation (IRT) method. We first construct the desired \emph{generalized linear-phase} spectrum:

\begin{equation}
H_{\mathrm{d}}(e^{j\omega}) = A_d(\omega)\,e^{j\psi_0}\,e^{-j\omega M/2},
\end{equation}
where $M$ is the filter order (thus we have $M+1$ coefficients) and $\psi_0$ is a constant phase offset that determines the type of symmetry:
$\psi_0 = 0$ (symmetric impulse response, Types I/II) or $\psi_0 = \pi/2$ (antisymmetric impulse response, Types III/IV).

{The IRT method is then applied to $H_{\mathrm{d}}(e^{j\omega})$ and the corresponding finite-length FIR impulse response becomes:}
\begin{eqnarray}
\label{eq:h}
& & h[n]={}\nonumber\\
& & {}=\!\left\{
\begin{array}{l l}
\!\! \frac{\displaystyle 1}{\displaystyle 2\pi}\displaystyle \!\! \int_{-\pi}^{\pi} \!\!\!\!\! A_d(\omega)\,e^{j\psi_0}\,e^{-j\omega M/2}\,e^{j\omega n}\,d\omega, 
& n = 0,1,\dots,M,\\[8pt]
\!\! 0, & \text{otherwise.}
\end{array}
\right.\nonumber\\
& & 
\end{eqnarray}
This impulse response is obtained by truncating the inverse Fourier transform of the shifted ideal response $H_{\mathrm{d}}(e^{j\omega})$ over the interval $n = 0, 1, \dots, M$. 
The resulting FIR filter $h[n]$ is \emph{causal} and exhibits \emph{Generalized Linear Phase (GLP)} behavior, with a constant group delay of $M/2$. The constant phase component $\psi_0$ specifies the corresponding symmetry type of $h[n]$.

\begin{figure}
\centering
\includegraphics[width=0.4\linewidth]{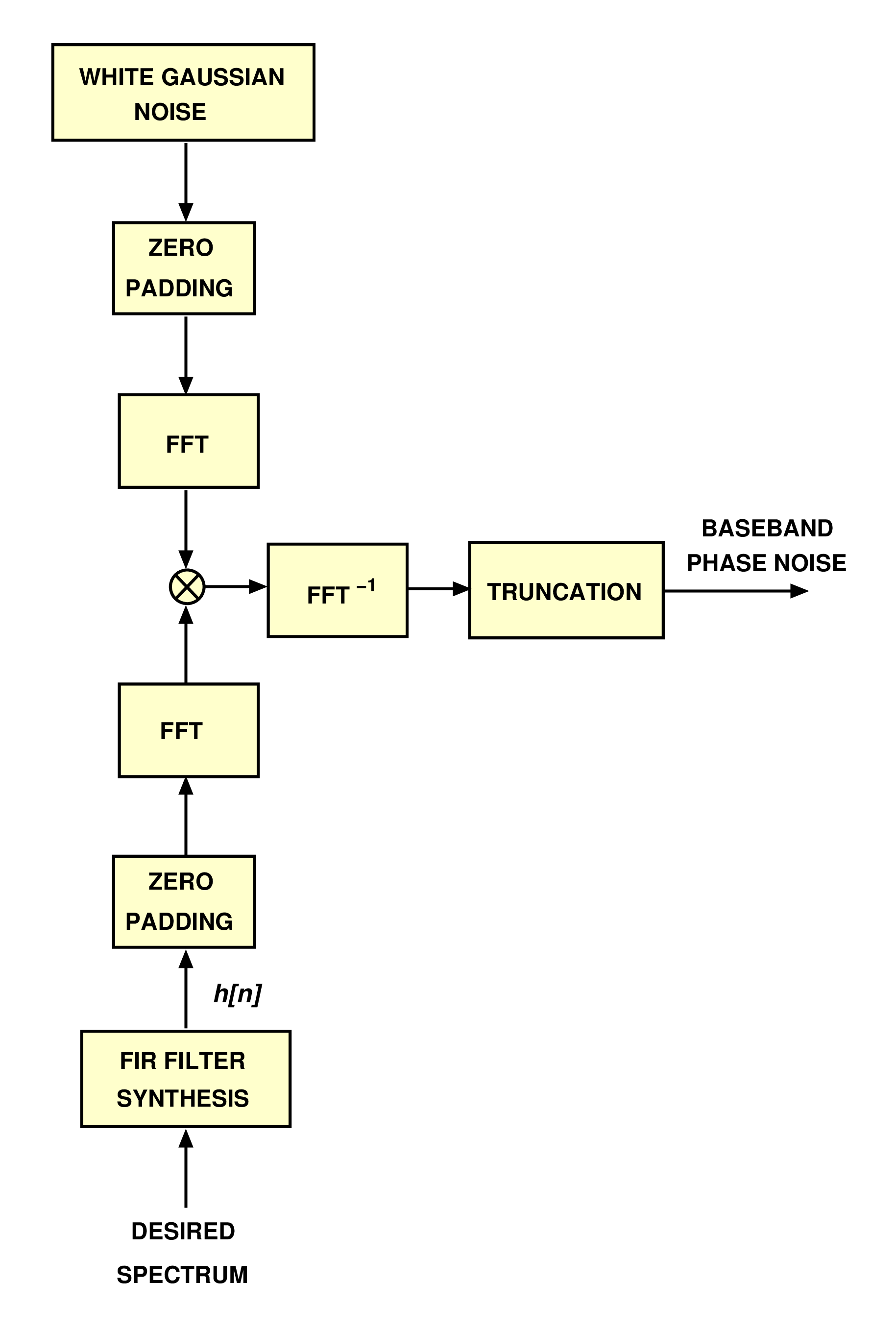}
\caption{Block diagram of the numerical procedure to generate baseband phase noise.}
\label{blocchi}
\end{figure}
 
{In our numerical procedure, the input white Gaussian sequence has unit variance, $\sigma_w^2=1$. Thus, if $S_{\phi,\mathrm{target}}(e^{j\omega})$ is the target PSD, we can set
$
A_d(\omega)
 =
 \sqrt{S_{\phi,\mathrm{target}}(e^{j\omega})}
$,
 $\psi_0=0$, and restrict the design to even values of $M$,
obtaining a Type-I FIR filter.
}
This choice is the safest and most suitable for the present application, since Type~I filters do not impose intrinsic zeros at either $\omega=0$ or $\omega=\pi$. As a result, they provide the most general linear-phase structure for approximating an arbitrary real amplitude response $A_d(\omega)$.

{The value of $M$ itself can be chosen according to the desired trade-off between approximation accuracy and computational complexity. A reasonable criterion is to select $M$ so that the normalized mean-square error $\mathrm{NMSE}(M)$ incurred by replacing the ideal impulse response $h_d[n]$ with the length-$(M+1)$ impulse response $h[n]$ is below a prescribed tolerance $\epsilon$. In the frequency domain, this condition is equivalent to finding the smallest even value of $M$ such that: }

\begin{equation}
\mathrm{NMSE}(M) = 
\frac{
\displaystyle \int_{-\pi}^{\pi}
\left|
H_{\mathrm{d}}(e^{j\omega}) - H(e^{j\omega})
\right|^2 d\omega
}{
\displaystyle \int_{-\pi}^{\pi}
\left|
H_{\mathrm{d}}(e^{j\omega})
\right|^2 d\omega
}
< \epsilon\, .
\end{equation}

{It is worth noticing that the IRT method minimizes the mean-square approximation error for a fixed filter order $M$. Thus, for a given number of coefficients, the truncated impulse response provides the least-squares optimal FIR approximation of the target frequency response.}

{In our simulations, we have always used a very large order, M=1000000, chosen by increasing it until the impulse response for the case with the narrowest bandwidth did not exhibit significant changes for further increases.}

The numerical procedure, shown in more detail in Fig.~\ref{blocchi}, allows us to obtain a phase noise process with the desired PSD, just by modeling the shape of the FIR filter frequency response. 
To accelerate the computation of the convolution in the time domain between the white Gaussian process and the impulse response of the filter, the Fast Fourier Transform (FFT) is leveraged to convert the operation into a multiplication in the frequency domain, performing proper zero-padding to ensure that the vector lengths are powers of 2 \cite{10.1007/978-3-031-71518-1_12}. Once the baseband phase noise values with specific amplitude and spectral characteristics are obtained, they are used directly as the phase $\phi_j$ at time $t_j$ of the carrier for the control pulses ($s(t_j)=\Re\{f(t_j) \exp[i(2\pi \nu t_j +\phi_j)]\}$, where $f(t_j)$ is the pulse envelope at $t_j$, and $\nu$ is the carrier frequency), in order to simulate their effect on qubit evolution. 

We use a sampling frequency of 1~GHz (corresponding to a time step of 1~ns), thus generating a baseband signal limited to a 500~MHz bandwidth, which ensures compliance with the Nyquist criterion.
{To achieve phase modulation of a 6~GHz carrier, the sampling frequency is subsequently increased by an integer factor via band-limited discrete-time resampling (Whittaker-Shannon sinc interpolation). Specifically, the temporal resolution is 1~ns at generation and becomes 83.3~ps after the interpolation}.

We performed the simulations of the qubit time evolution with Qiskit-Dynamics~\cite{qiskit}, an open-source software which solves the Hamiltonian of the interaction between the electromagnetic field and the qubit. The software numerically integrates the time-dependent Schr\"odinger equation to compute the evolution of a qubit under the effect of a control Hamiltonian. {Specifically, we configured Qiskit Dynamics (version 0.4.4) with the JAX backend in double precision and performed the integration using the \textit{jax\_odeint} solver. To ensure numerical accuracy and proper resolution of the control waveforms, both the absolute and relative tolerances were set to $10^{-8}$ ($\text{atol} = \text{rtol} = 10^{-8}$), and the maximum internal step size of the solver was strictly constrained to the pulse sampling interval, $\text{max\_step} = dt = 83.3\text{ ps}$.
This allows us to calculate} the complete temporal evolution of the qubit state starting from a defined initial condition and driven by control pulses with user-defined parameters. From the solver output, we obtain the temporal evolution of the qubit state vector, which enables us to evaluate the fidelity between the final state and the ideal state predicted on the basis of the noise-free evolution, thus quantifying the effect of phase noise on the accuracy of qubit control.

\section*{Results and Discussion}\label{sec_two}


\subsection*{Contribution of individual noise spectral components}\label{subsec_one}

\begin{figure}
\centering
\includegraphics[width=0.5\linewidth]{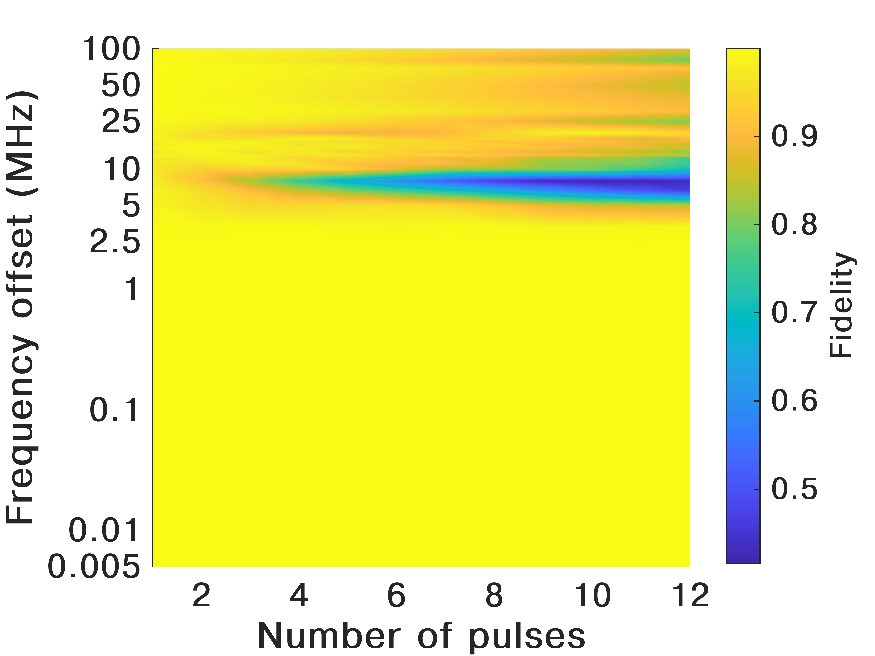}
\caption{Fidelity as a function of the frequency offset and of the number of 50~ns $\pi_x$-pulses applied with constant amplitude of $-60$~dBc/Hz. The qubit is initially prepared in $|0\rangle$. 
{The presence of a few fidelity values slightly below 0.5 is due to fluctuations resulting from the limited number of averages.}}
\label{gauss_50ns}
\end{figure}

Our initial objective has been that of understanding the contribution of the different frequency components in the phase noise spectrum around the carrier. We thus considered a phase noise PSD with a Gaussian shape, a bandwidth  of 2.1~kHz and a constant amplitude of $-60$~dBc/Hz, centered on several frequencies. This allowed us to investigate the impact of the individual phase noise frequency components, varying the center frequencies of the filters while maintaining a constant noise power. The performed tests are based on the previously described approach, with sequences of applied $\pi_x$ pulses. At the beginning of the first simulation, the qubit state is set to $|0\rangle$ and a sequence of 12 consecutive 50~ns rectangular control pulses, spaced 20~ns apart (implementing 12 consecutive NOT gates) is applied. The amplitude of the pulses is initially tuned to achieve the ideal scenario of near-perfect fidelity, in the absence of noise, i.e., in such a way as to make each 50~ns pulse an ideal $\pi_x$-pulse. Finally, the generated phase noise values are applied to the phase of the control pulses. The fidelity of the states resulting after each pulse is then evaluated to quantify the impact of phase noise over time. 
In Fig.~\ref{gauss_50ns}, the results of this first simulation are shown. In particular, fidelity values are reported as a function of the frequency offset of each Gaussian PSD from the carrier and of the number of {applied 50~ns $\pi_x$-pulses.} The noise amplitude has been chosen to be unrealistically large, in order to obtain a significant loss of fidelity, so that a reduced number of averages is sufficient to get a good estimate.
We observe that the largest contributions to the loss of fidelity occur around the Rabi frequency, which in this case is $1/(2 \cdot 50$~ns)$=10$~MHz.

\begin{figure}
\centering
\includegraphics[width=0.9\linewidth]{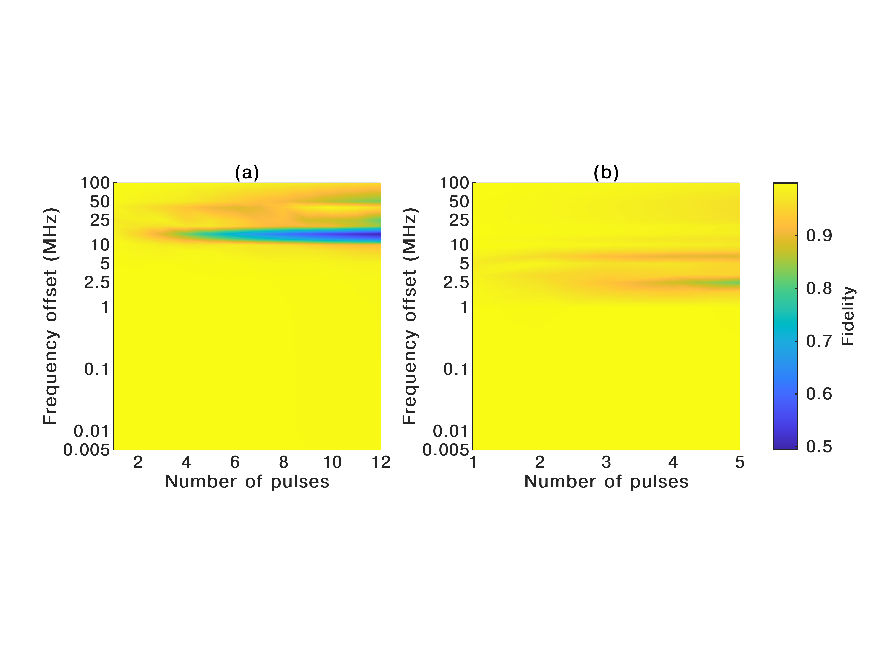}\hfill
\caption{Fidelity as a function of the number of applied $\pi_x$-pulses, with a constant amplitude of $-60$~dBc/Hz, and of the central frequency of the Gaussian filter used to model phase noise. 
In both panels the state of the qubit is initially prepared in $|0\rangle$. The left panel (a) shows the results for a sequence of 25~ns pulses separated by 10~ns, while the right panel (b) refers to 150~ns pulses with an inter-pulse interval of 60~ns.}
\label{25ns150nsgaussiani_assez}
\end{figure}

We have then run other simulations with different choices of the number of pulses, their duration and their separation, in order to better investigate the role of the various parameters. In Fig.~\ref{25ns150nsgaussiani_assez}, the fidelity results for two tests are reported. The data shown in the left panel are for a sequence of 12 pulses 25~ns long, separated by 10~ns intervals, while those in the right panel are for a sequence of 10 pulses 150~ns long, separated by 60~ns intervals. We observe that, as in the previously discussed example, the largest effect on fidelity is due to components close to the Rabi frequencies of the individual experiments, $ 20$~MHz and $3.33$~MHz, respectively. 

After analyzing the case in which the qubit final state is along the $z$-axis, we now consider operations involving final states on the equatorial plane of the Bloch sphere (although they cannot be measured directly). In this situation, we have to take into account the fact that the phase of the qubit is always defined relative to the instantaneous phase of the reference clock used for the control. When no resonant drive is applied, the qubit does not physically interact with the reference clock fluctuations; however, the control reference frame evolves affected by the fluctuations due to phase noise, leading to a relative shift with respect to that of the qubit. 

\begin{figure}
\centering
\includegraphics[width=0.9\linewidth]{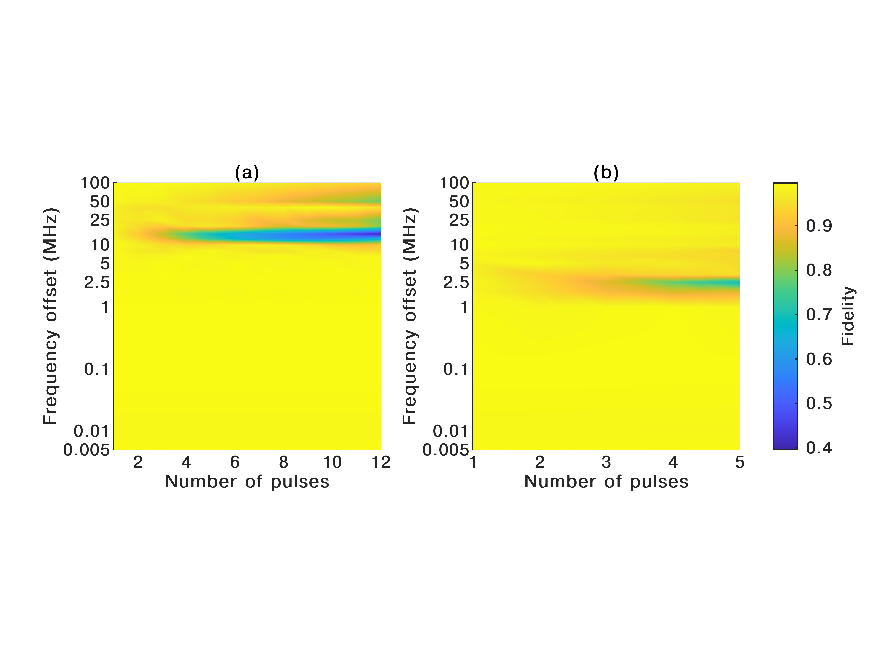}\hfill
\caption{Fidelity as a function of the number of applied $\pi_x$-pulses, with a constant amplitude of -60~dBc/Hz, and of the central frequency of the Gaussian filter used to model phase noise. 
In both panels the qubit is initially prepared on the equatorial plane of the Bloch sphere, along the $y$-axis. The left panel (a) shows the results for a sequence of 25~ns pulses separated by 10~ns, while the right panel (b) refers to 150~ns pulses with an inter-pulse interval of 60~ns.
{The presence of a few fidelity values slightly below 0.5 is due to fluctuations resulting from the limited number of averages.}}
\label{25ns150nsgaussiani_assey}
\end{figure}

Since fidelity must be evaluated with reference to the control frame (because all subsequent operations will be referred to such a frame), the target state has to be expressed in the same basis. Therefore, for each fidelity evaluation we add the instantaneous reference clock phase deviation to the azimuthal angle of the target state, thus aligning the reference for the evaluation of fidelity with the frame of the local oscillator at the measurement time.
This procedure allows us to correctly account for the instantaneous phase deviation of the reference oscillator when evaluating operations on states in the equatorial plane.
We then applied the same pulse sequences discussed above, with the same noise power, to a qubit with an initial state on the equatorial plane along the $y$-axis (Fig.~\ref{25ns150nsgaussiani_assey}), obtaining a somewhat increased degradation of fidelity. This was expected, because of the presence of the additional contribution due to the fluctuations around the $z$-axis. 

\begin{figure}
\centering
\includegraphics[width=0.5\linewidth]{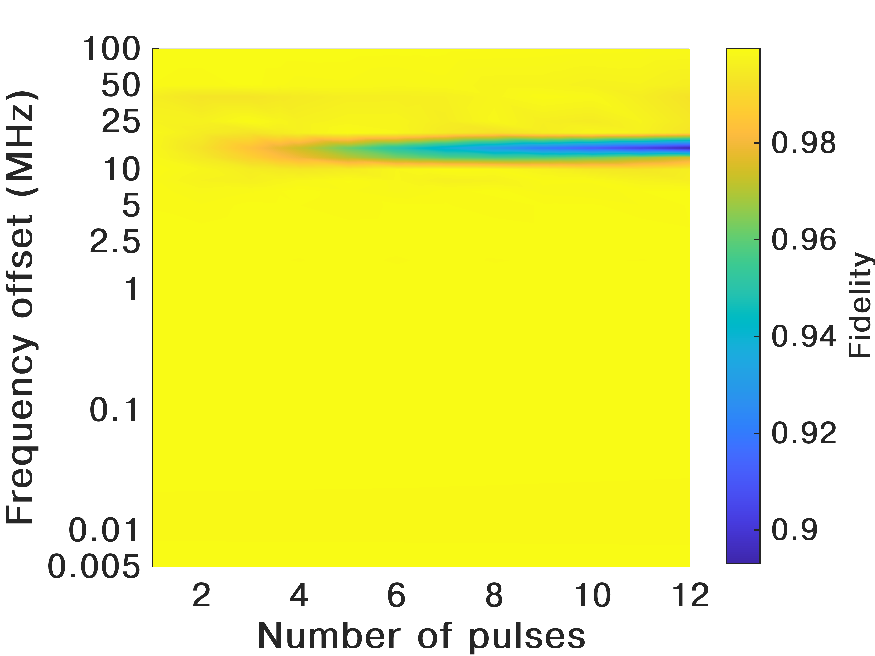}
\caption{Fidelity as a function of frequency offset and of the number of 25~ns $\pi_x$-pulses applied with a constant amplitude of $-70.5$~dBc/Hz. The qubit is initially prepared in $|0\rangle$.}
\label{scalfact0p09}
\end{figure}

As an additional test, we performed the same simulation with a sequence of 12 $\pi_x$-pulses of 25~ns duration and the qubit initially prepared in $|0\rangle$, but this time assuming a constant noise amplitude of $-70.5$~dBc/Hz, which resulted in a minimum fidelity of 0.9 (Fig.~\ref{scalfact0p09}).

\begin{figure}
\centering
\includegraphics[width=0.9\linewidth]{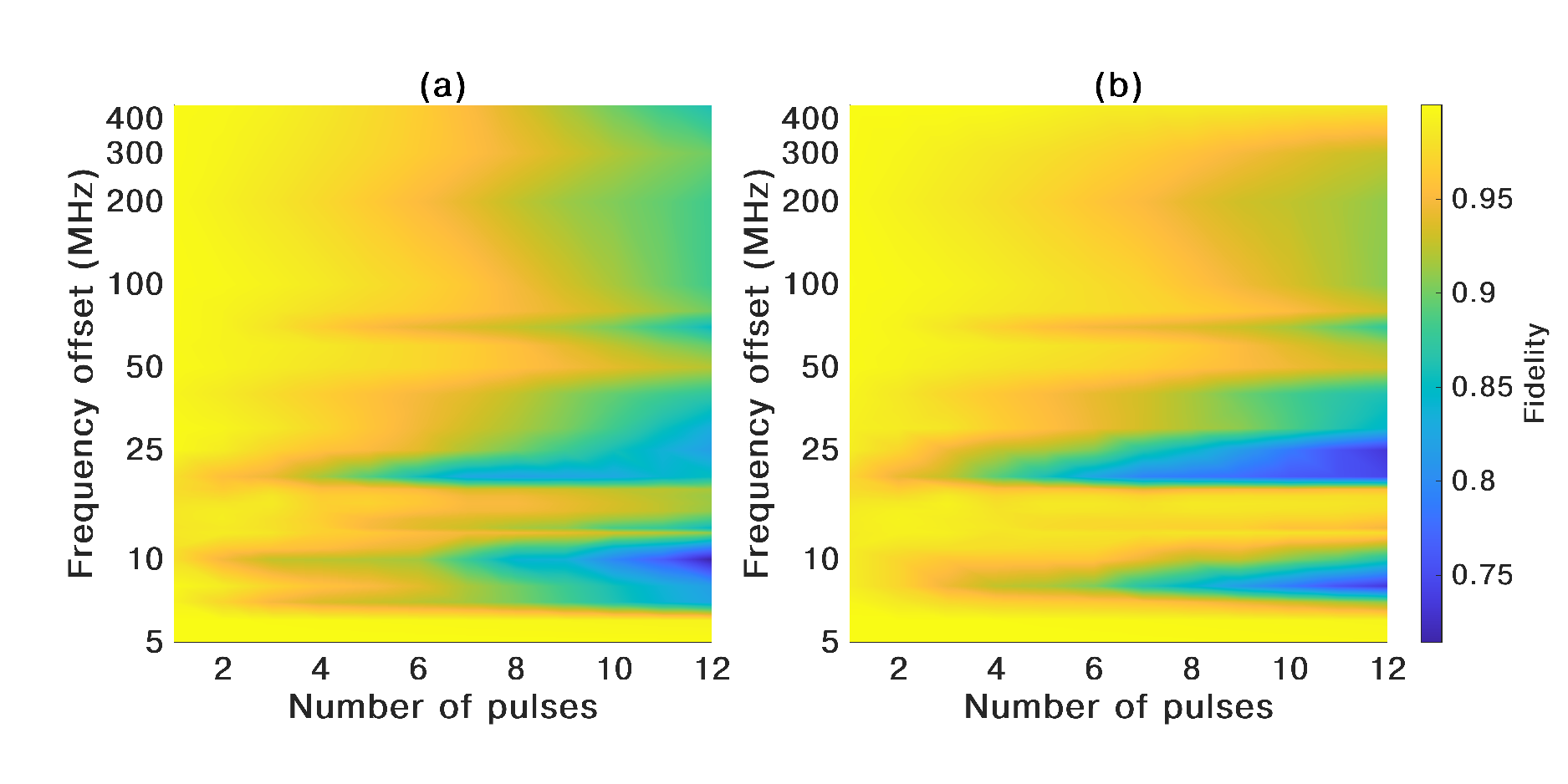}
\caption{{{Fidelity as a function of the frequency offset and of the number of 50~ns $\pi_x$-pulses (rectangular for panel (a) and DRAG for panel (b)) applied with a constant amplitude of $-80$~dBc/Hz and a bandwidth of the Gaussian FIR filter of 2.1~MHz. The qubit is initially prepared in $|0\rangle$}.}}
\label{25ns_6MHz}
\end{figure}

{In order to understand how the pulse shape influences the effect of phase noise on fidelity, we have performed a simulation both for rectangular (panel (a)) and DRAG (Derivative Removal by Adiabatic Gate) (panel (b)) pulses with a duration of 50~ns. Phase noise has a  Gaussian shaped  power spectral density with a bandwidth of 2.1~MHz (Fig.~\ref{25ns_6MHz}) and an amplitude of $-80$~dBc/Hz. Considering the filter bandwidth, the center frequency of the Gaussian noise spectrum is swept from 5~MHz to 450~MHz. It is apparent that, once again, the maximum effect is close to the Rabi frequency and that high frequency components of the noise spectrum have a reduced effect (which in practice becomes negligible because in the case of a realistic oscillator they have a much lower amplitude with respect to the low frequency components, as discussed in the following). We notice that with DRAG pulses we have a reduced fidelity deterioration at offsets near the Rabi frequency, while at higher offsets {there is no significant difference with respect to the case of rectangular pulses.}} The noise power level has been lowered with respect to the previous case, because otherwise the total noise power associated with the Gaussian spectrum would be much larger (due to the increased bandwidth), leading to an extremely low fidelity. 

{To provide an overview of the results achieved thus far, the simulation data for the rectangular pulse sequences are summarized in Table \ref{summary_rect}. This allows a direct comparison between different choices of parameters. Specifically, we considered a selection of pulse numbers and durations, initial states, and noise bandwidths.}

\begin{table}[ht]
    \centering
    \footnotesize 
    \setlength{\tabcolsep}{4pt} 
    \begin{tabular}{l c c c c c c c c}
        \toprule
        \textbf{Fig.} & 
        \makecell{\textbf{Pulse} \\ \textbf{Num.}} &
        \makecell{\textbf{Initial} \\ \textbf{State}} & 
        \makecell{\textbf{$\Omega_R$} \\ \textbf{(MHz)}} & 
        \makecell{\textbf{Ampl.} \\ \textbf{(dBc/Hz)}} & 
        \makecell{\textbf{Band} \\ \textbf{(kHz)}} &
        \makecell{\textbf{Fid.} \\ ($f<\Omega_R$)} & 
        \makecell{\textbf{Fid.} \\ ($f\simeq\Omega_R$)} & 
        \makecell{\textbf{Fid.} \\ ($f>\Omega_R$)} \\
        \midrule
        2   & 12 & $|0\rangle$ & $10$ & -60 & 2.1 &$\sim 1.0$ & $\sim0.5$ & $>0.9$ \\
        3.a & 12 & $|0\rangle$ & $20$ & -60 & 2.1 & $\sim 1.0$ & $\sim0.5$ & $>0.8$ \\
        3.b & 10 & $|0\rangle$ & $3.33$ & -60 & 2.1 & $\sim 1.0$ & $\sim0.8$ & $>0.9$ \\
        4.a  & 12 &  along $-y$-axis & $20$ & -60 & 2.1 & $\sim 1.0$ & $\sim0.5$ & $>0.75$ \\
        4.b  & 10 & along $-y$-axis & $3.33$ & -60 & 2.1 & $\sim 1.0$ & $\sim0.75$ & $>0.9$ \\
        5 & 12 & $|0\rangle$ & $20$ & -70.5 & 2.1 & $\sim 1.0$ & $\sim0.89$ & $>0.98$ \\
        6.a & 12 & $|0\rangle$ & $10$ & -80 & 2100 & $\sim 1.0$ & $\sim0.75$ & $>0.85$ \\
        \bottomrule
    \end{tabular}
    \caption{{Summary of simulation results for the case of rectangular pulses presented in the ``Analytical approximation, interpretation and numerical validation'' subsection.}}
    \label{summary_rect}
\end{table}

\subsection*{Analytical approximation, interpretation, and numerical validation}\label{subsec_two}

From the results we have presented so far, it is apparent that {the largest relative contribution} of phase noise to the loss of fidelity is from frequency components that are at an offset, relative to the qubit resonant frequency, close to the Rabi frequency, which is intuitively reasonable, because the period of the Rabi frequency is the basic timescale of qubit evolution.

On the other hand, it is less intuitively clear why components with high frequency displacement with respect to the resonant frequency could affect the qubit evolution at all, since the qubit, which is intrinsically a resonator, should be substantially insensitive to them. In the following, we will discuss this aspect in detail.

Since the level of phase noise affecting the reference clock of the qubit control equipment is usually very low, modulation of the carrier by phase noise can be approximated, up to second order in $\theta(t)$, as:
\begin{eqnarray}\label{modulazioni}
& & \cos[\omega_0 t + \theta(t)] = \cos(\omega_0 t) \cos[\theta(t)] - \sin(\omega_0 t) \sin[\theta (t)] \simeq {} \nonumber\\
& & {} \simeq \cos(\omega_0 t) [1 -\theta(t)^2/2]- \sin(\omega_0 t)\theta(t)\, ,
\end{eqnarray}
where $\omega_0$ is the control pulse carrier frequency and $\theta(t)$ is the instantaneous (random) phase deviation. Thus, while from the second term on the right hand side we obtain a double-sideband modulation (with suppressed carrier) of the phase noise spectrum around the carrier frequency, from the first term we obtain a residual amplitude modulation of the carrier. 

Let us initially discuss the contribution of the second term, which leads to a direct
translation of the noise spectrum around the carrier: 
thus, the coupling between a noise spectral component and the qubit is analogous to the coupling between a pure sine control signal with the same displacement with respect to the resonance frequency and the qubit. Such a coupling decreases as displacement increases, until it practically vanishes. 

\begin{figure}  
\centering
\includegraphics[width=0.5\linewidth]{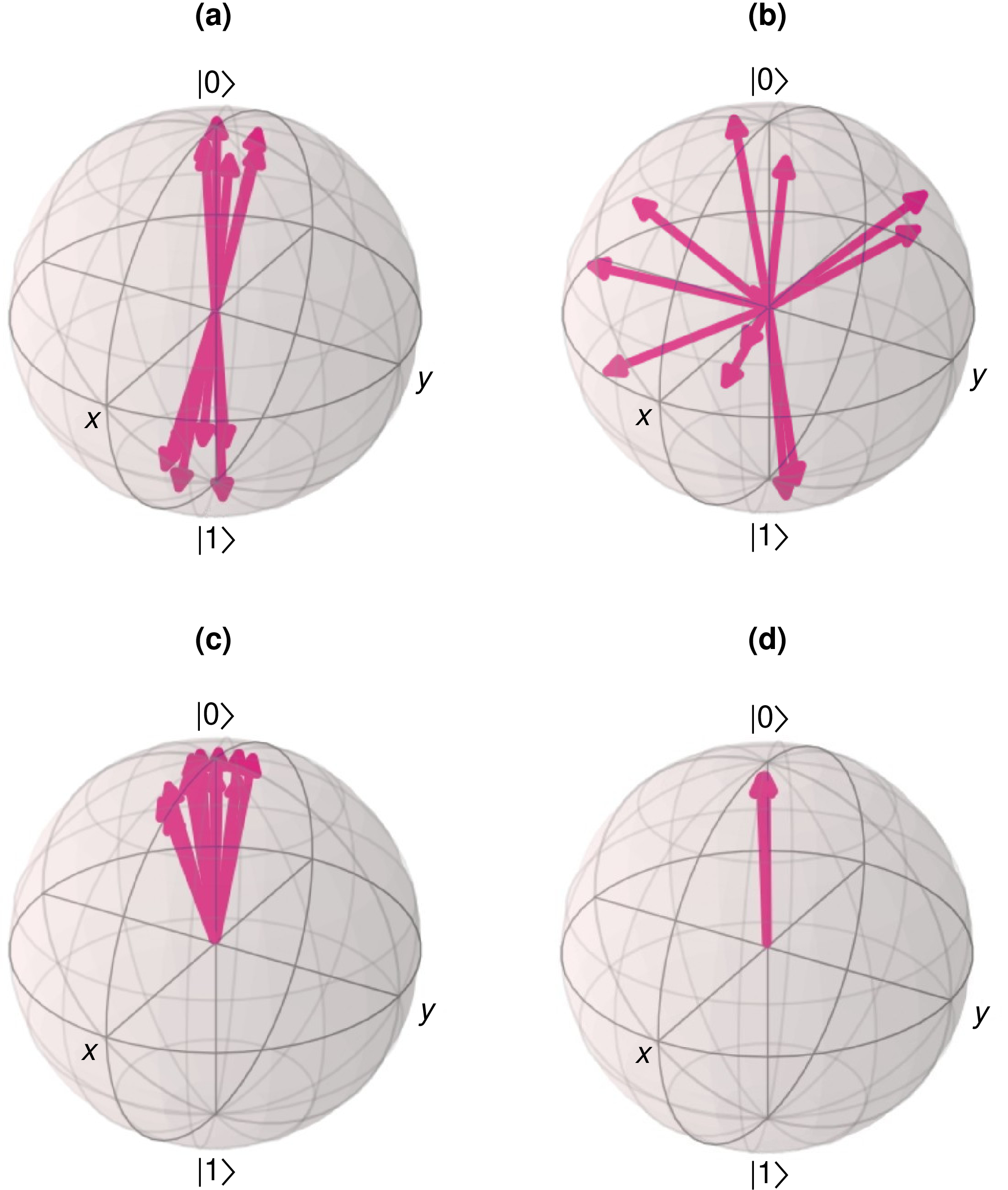}
\caption{Bloch sphere representation of a qubit state driven by a pulse with increasing detuning from the resonance frequency. For a detuning of 1~MHz (a), we have transitions of the qubit state from $ | 1\rangle$ to $ | 0\rangle$, indicating an efficient population transfer; for a detuning of 10~MHz (b) coupling is still present but the states become scattered all around the Bloch sphere; for a detuning of 50~MHz (c), the state vector mainly precesses around the $z$-axis; for a detuning of 400~MHz (d), the state vector remains unchanged because coupling is suppressed. This evolution of the polar angle is mainly due to the impact of the amplitude modulation effect.}
\label{testsenzarumore}
\end{figure}

This can be shown with a simple simulation: we have applied to the qubit the 12-pulse sequence previously used to obtain the results in Fig.~\ref{gauss_50ns}, but in this case in the absence of noise. The outcomes of four tests, each with increasing detuning of the pulse carrier from the resonance frequency, are presented in Fig.~\ref{testsenzarumore}. As detuning increases, it is apparent that the pulses are progressively less effective in driving the expected state transitions or even any perturbation of the state at all. Specifically, with a detuning of 1~MHz (Fig.~\ref{testsenzarumore}(a)) state inversion with population transfer is still obtained, while for $10$~MHz (Fig.~\ref{testsenzarumore}(b)) significant coupling to the qubit is still present, but the final state wanders around the Bloch sphere. However, with a 50~MHz detuning (Fig.~\ref{testsenzarumore}(c)), coupling is significantly reduced, with oscillations mainly due to precession around the $z$-axis. Finally, with a 400~MHz detuning, the initial state is completely preserved, because coupling vanishes.

{High frequency components do, however, provide a nonvanishing contribution via the residual amplitude modulation due to the term $\cos(\omega_0 t) [1 -\theta(t)^2/2]$ in Eq.~(\ref{modulazioni}): the squared term $\theta(t)^2$ introduces a negative zero frequency component which reduces the carrier amplitude, thereby decreasing the amount of rotation around the $x$-axis. It also introduces two side bands, but they have a very limited effect because they are shifted to higher frequencies and they have a second order amplitude.}

In the case of final states along the $z$-axis of the Bloch sphere, fidelity
is not affected by the phase difference between the reference clock frame and the qubit frame (in such a case azimuthal rotations are irrelevant). Such a difference has, indeed, an effect in the case of a final state on the equatorial plane.

This effect must always be included in the results of our simulations (which focus on the direct interaction between the control field and the qubit) whenever the final state is not aligned with $z$. 

{In order to quantify the impact of the two terms described above, we present two illustrative examples. We consider a sequence of twelve rectangular 50~ns $\pi_x$-pulses, affected by a Gaussian phase-noise power spectral density with a bandwidth of 2.1~MHz and a level of $-80$~dBc/Hz. Two offset frequencies are considered: 12.5~MHz and 300~MHz. {To isolate the individual effects of the residual amplitude modulation and of the quadrature DSB modulation, their contributions are applied to the $I$ and $Q$ components of the pulses, respectively, according to the analytical expressions of Eq.~(\ref{modulazioni})}. The results summarized in Table \ref{IQcontr} reveal a clear spectral distinction between the two error contributions. The contribution of the DSB modulation term to the loss of fidelity exhibits a strong frequency dependence; specifically, for offset values close to the Rabi frequency, it is the primary source of error, whereas its contribution decreases significantly and becomes negligible at high offset frequencies, such as 300~MHz. In contrast, the contribution related to the residual amplitude modulation term behaves as a {substantially frequency-independent term {(due to the mentioned carrier suppression)}}: although it is much smaller than that of the DSB term for offsets of approximately the Rabi frequency, it persists uniformly across the entire spectrum.}

\begin{table}[htbp]
\centering
\setlength{\tabcolsep}{12pt} 
\begin{tabular}{l S[table-format=1.5] S[table-format=1.5] S[table-format=1.5]}
\toprule
\textbf{Frequency offset} & {\textbf{Residual AM}} & {\textbf{DSB}} & {\textbf{Residual AM and DSB}} \\
\midrule
12.5~MHz & 0.92697 & 0.54215 & 0.52979 \\
300~MHz  & 0.92559 & 0.99997 & 0.92544 \\
\bottomrule
\end{tabular}
\caption{{Quantitative comparison of the individual and combined effects of residual AM and DSB modulations on fidelity after 12 rectangular 50~ns $\pi_x$-pulses.}}
\label{IQcontr}
\end{table}

Let us now see a simple example for a state on the equatorial plane, discussing one effect at a time: the effect on the polar angle, mainly resulting from the amplitude modulation by noise of the pulse carrier, the effect on the azimuthal angle of the double-sideband (DSB) modulation, and, finally, the additional effect on the azimuthal angle of the instantaneous phase shift between the qubit frame and the local oscillator frame.

\begin{figure}
\centering
\includegraphics[width=0.5\linewidth]{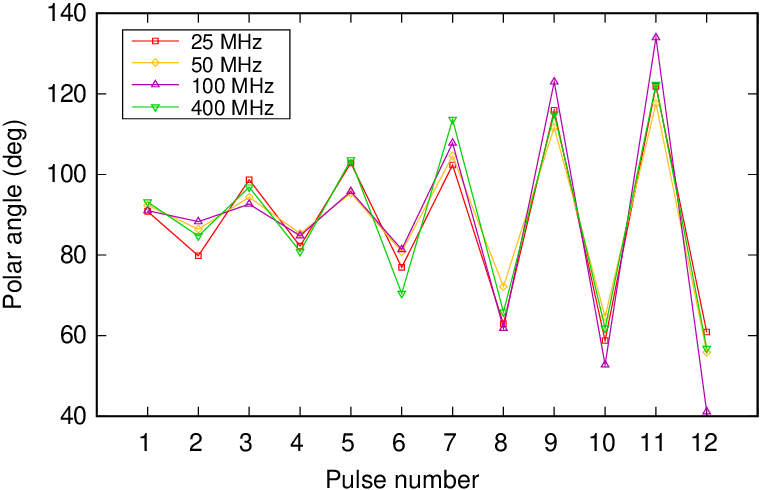}
\caption{Evolution of the polar angle. The results correspond to a qubit initially prepared along the negative $y$–axis and driven by a sequence of twelve 25~ns $\pi_x$–pulses, in the presence of phase noise generated through Gaussian FIR filters with different center frequencies and with bandwidth of 4.2~MHz.}
\label{polari}
\end{figure}

We start with a state on the equatorial plane, aligned with the $y$-axis, and apply $\pi_x$ pulses causing the rotation around the $x$-axis: the residual amplitude modulation will act only upon the polar angle, since it is varying the amplitude of the control signal and therefore the overall rotation angle. In Fig.~\ref{polari}, we report the evolution of the polar angle during a sequence of 12 $\pi$-pulses with a duration of 25~ns (and a separation between pulses of 10~ns) for a qubit initial state along the negative $y$-axis. In this case, we generated phase noise sequences through a Gaussian FIR filter with varying center frequencies and a bandwidth of 4.2~MHz. {We notice that the effect on the phase angle is essentially independent of frequency. There is an 
oscillatory behavior because the state oscillates between the negative and the positive $y$ subspace: let us assume that after the first pulse the state has rotated by an angle slightly less than 180~degrees and therefore the polar angle is slightly more than 90~degrees, let us say $90+\alpha$~degrees, the second pulse will cause a rotation again of slightly less than 
180~degrees, which will lead to a polar angle of $90-\beta$~degrees, with $\beta > \alpha$, and so forth, therefore an oscillatory behavior with growing amplitude will be observed. The reason why the rotation is always slightly less than 180~degrees is that phase noise applied to an in-phase signal determines a partition of the pulse power between the in-phase component and the quadrature component. The in-phase component will always be smaller than the one without noise, as can be deduced from Eq.~(\ref{modulazioni}), where $\cos(\omega_0 t)$ is multiplied by $\cos(\theta)$, which is always less than 1. The growth of the phase error could in practice be counteracted by slightly increasing the carrier amplitude.}   

\begin{figure}  
\centering
\includegraphics[width=0.5\linewidth]{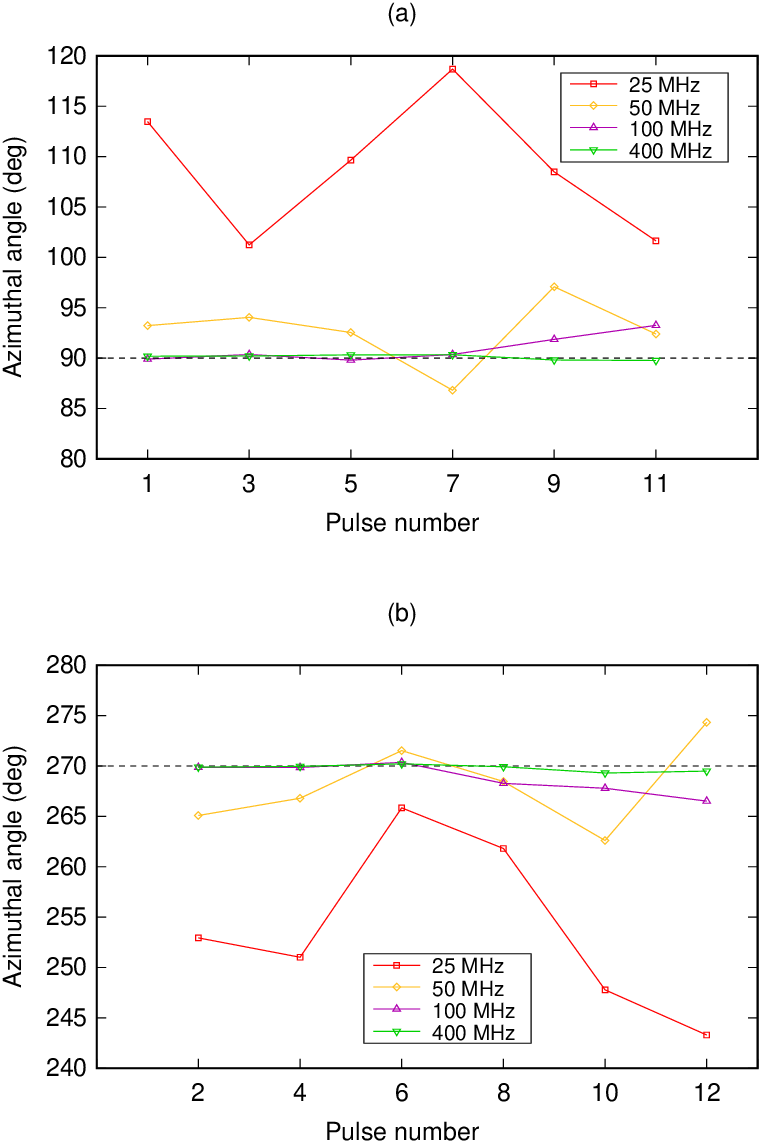}
\caption{Evolution of the azimuthal angle, taking into account only the output of the Qiskit-Dynamics simulation (without including the relative phase between the control reference clock frame and the qubit frame).  The results are for a qubit initially prepared along the negative $y$–axis and driven by a sequence of twelve 25~ns $\pi_x$–pulses, in the presence of phase noise with Gaussian power spectral densities with different center frequencies and with a bandwidth of 4.2~MHz.}
\label{angoli_azimutali1}
\end{figure}

In Fig.~\ref{angoli_azimutali1} we show the behavior of the component of the azimuthal angle directly obtained from the simulations for (a) odd and (b) even pulses. We see that in this case the effect, now mainly due to the quadrature component, decreases as we raise the frequency and get farther from the Rabi frequency, until it substantially disappears at 400~MHz. 

\begin{figure}  
\centering
\includegraphics[width=0.5\linewidth]{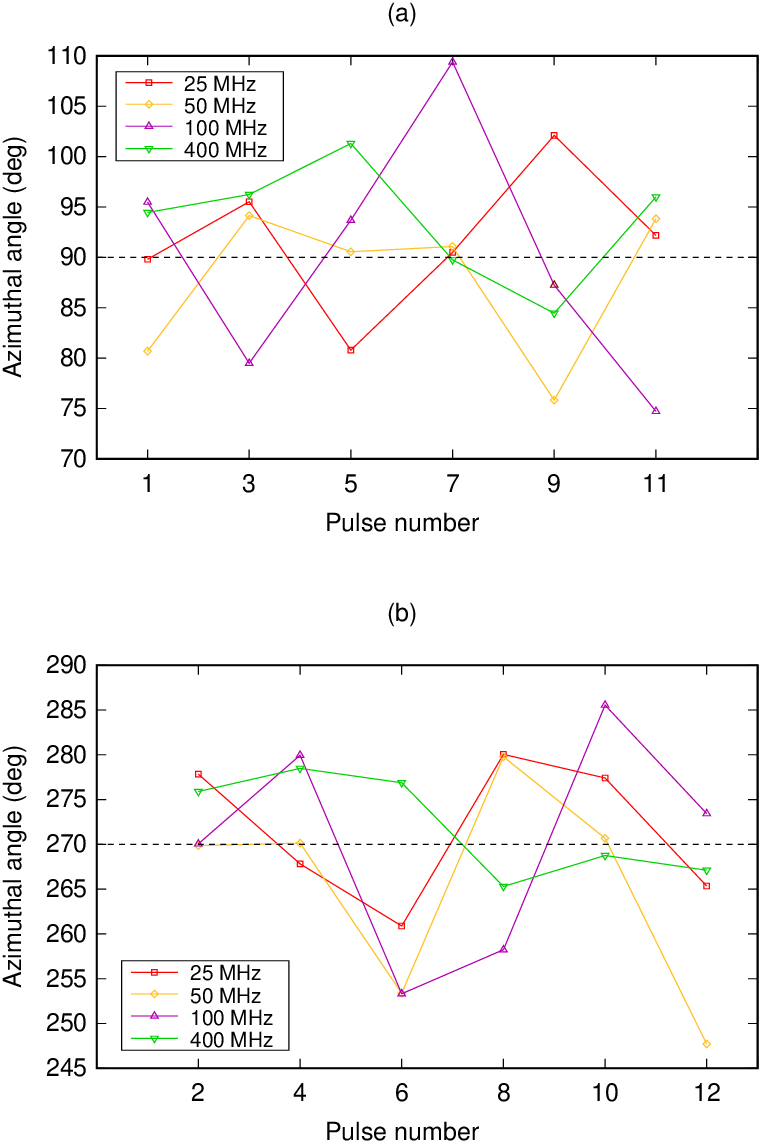}
\caption{Evolution of the azimuthal angle, taking into account only the effect of the instantaneous phase deviation of the control signal. The results are for a qubit initially prepared along the negative $y$–axis and driven by a sequence of twelve 25~ns $\pi_x$–pulses, in the presence of phase noise with Gaussian power spectral densities with different center frequencies and with a bandwidth of 4.2~MHz.}
\label{angoli_azimutali2}
\end{figure}

In Fig.~\ref{angoli_azimutali2} we report a plot of the component of the azimuthal angle due to the instantaneous value of the phase noise in the control signal, for the same pulse sequence as in the case of Fig.~\ref{polari}.
Panel (a) is for the odd pulse numbers and panel (b) for the even pulse numbers. The behavior is random because this is just the sequence of the random phase samples plus 90 (odd pulses) or 270 (even pulses) degrees. Contrary to previously discussed effects, this does not exhibit any accumulation in time.

\subsection*{Contribution of realistic and synthetic noise PSDs}\label{subsec_three}

In this Section, we apply our numerical technique to the study of the effect of the phase noise due to two different sources: an actual microwave oscillator~\cite{papi2mtc} that we developed for the up- and down-conversion of control pulses and readout signals,
and two synthetic microwave sources with a purposely designed noise PSD.

The former simulation has been performed just to verify the expected performance of our qubit control system in terms of maximum achievable fidelity, while the latter simulation has been devised to provide further evidence that the high-frequency components of phase noise {do not have a prevalent effect on fidelity, as already discussed in the previous sections.}

\begin{figure}
\centering
\includegraphics[width=0.5\linewidth]{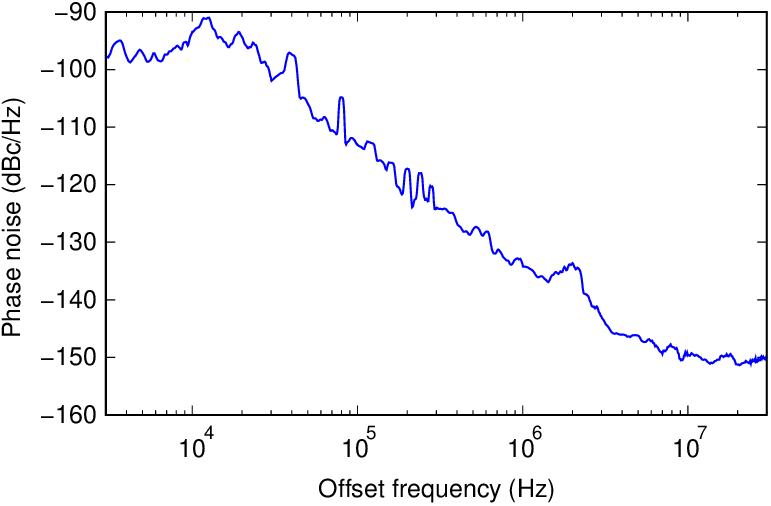}
\caption{PSD of the phase noise as a function of the frequency offset for our oscillator for a carrier at 3.4~GHz.}
\label{phasenoiseadf5355}
\end{figure}

We measured the phase noise of an improved version of the oscillator of Ref.~\cite{papi2mtc} with a Rohde \& Schwarz FSW spectrum analyzer, and we report the results in Fig.~\ref{phasenoiseadf5355}, for a carrier at 3.4~GHz. We notice that, while above about 20~kHz the PSD decays with a slope that is approximately $1/f^2$, below 20~kHz we have a peak preceded by a plateau: this is consistent with the action of the phase-locked loop, which tends to stabilize the phase of the voltage controlled oscillator (VCO) within the loop bandwidth (assuming that the crystal-based reference oscillator has a lower phase noise compared to the VCO). 

The application of phase noise with a PSD equal to that in Fig.~\ref{phasenoiseadf5355} to a sequence of ten 150~ns $\pi_x$-pulses leads to a very small reduction in fidelity, of the order of $10^{-5}$, and therefore close to the numerical accuracy of our method. 

In this case the noise components at low displacements from the carrier (and with the largest amplitude) have a small effect because the total duration of the pulse sequence (2~$\mu$s) and therefore the observation time are much shorter than the reciprocal of such frequencies.

In order to better understand the role played by the various frequency components, we ran a simulation analogous to those previously discussed with Gaussian shaped filters. This time, however, we used the data from Fig.~\ref{phasenoiseadf5355} to assign relative amplitudes for the FIR filters at the various frequencies.

{Since we use a Gaussian filter with a bandwidth of 2.1~kHz (the choice of a narrow bandwidth is due to the fact that we start the evaluation of the effect of phase noise from an offset of just 10~kHz) we need to raise the noise amplitude at each frequency by about 75~dB.} 

\begin{figure}
\centering
\includegraphics[width=0.5\linewidth]{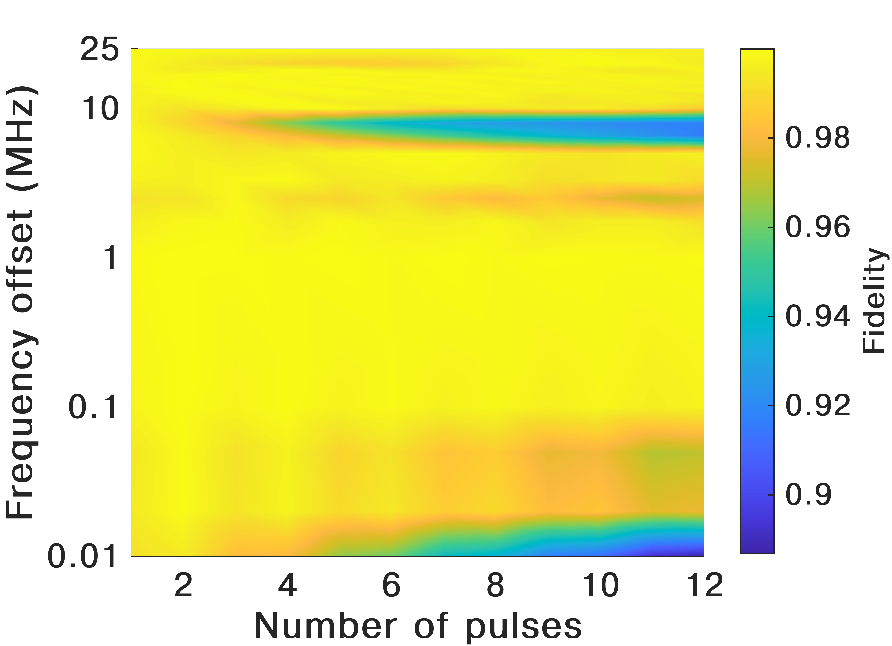}
\caption{Fidelity as a function of the frequency offset and of the number of applied 50~ns $\pi_x$-pulses, with phase noise amplitudes scaled proportionally to the amplitude of the actual phase noise spectrum of a microwave oscillator.}
\label{gauss_pesati}
\end{figure}

We have considered a sequence of twelve 50~ns $\pi_x$-pulses separated by 20~ns intervals. The results are reported in Fig.~\ref{gauss_pesati}, and, as expected, the components close to the Rabi frequency have a predominant effect. 
However, there is a significant contribution also at low frequencies, resulting from the very large value of the PSD in this frequency range.

{The calculation we just presented serves the purpose of showing the relative effect of each component of the noise spectrum, while, in order to evaluate the actual effect of the PSD reported in Fig.~\ref{phasenoiseadf5355} (with the original amplitude), we implemented a FIR filter yielding the complete PSD spectrum. For the same pulse sequence as in Fig.~\ref{gauss_pesati} we obtained a fidelity of 0.99998, while for a sequence of ten 150~ns $\pi_x$-pulses we computed a fidelity of 0.997.}

\begin{figure}
\centering
\includegraphics[width=0.7\linewidth]{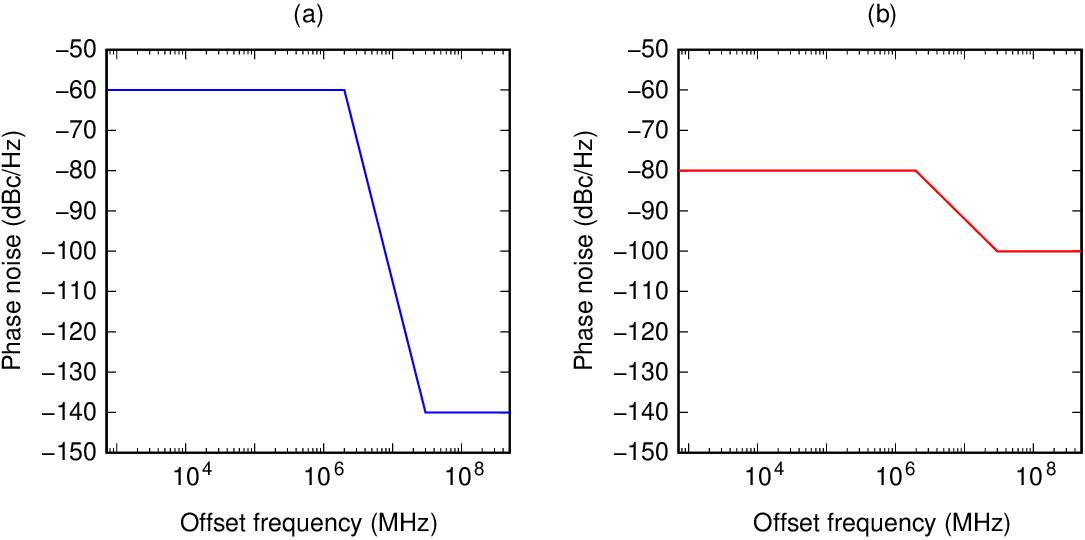}
\caption{Synthesized phase noise PSD to study the effect on the fidelity of the frequency components:
(a) the noise level is set high at low frequencies (-60~dBc/Hz) and very low at high-frequencies (-140~dBc/Hz); (b) the high-frequency noise components are increased (-100~dBc/Hz) while the low-frequency components are reduced (-80~dBc/Hz).
}
\label{psd_confronto_rumori}
\end{figure}

\begin{figure}
\centering
\includegraphics[width=0.55\linewidth]{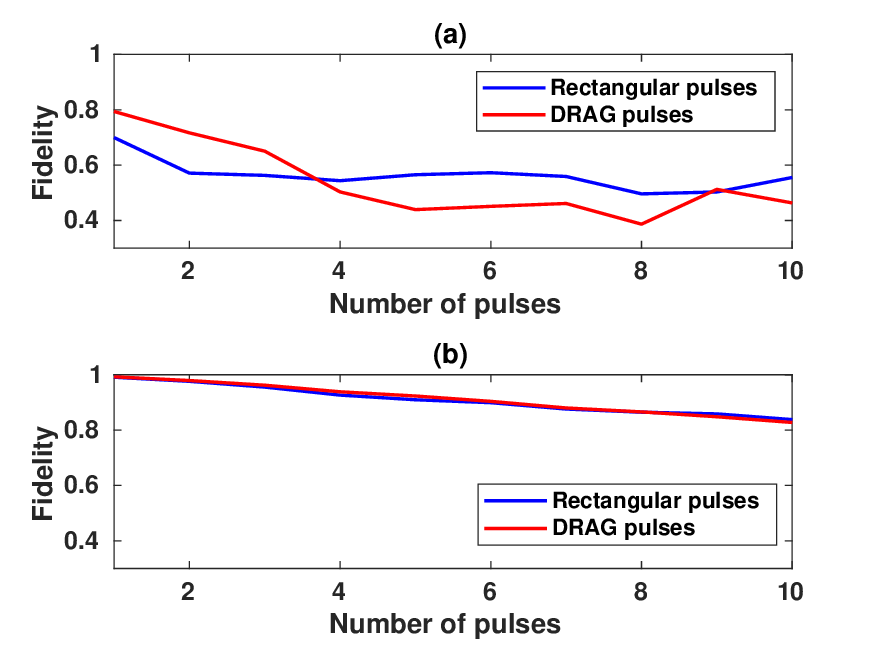}
\caption{Fidelity as a function of the number of 150~ns applied $\pi_x$-pulses, with the phase noise values from the model described in Fig.~\ref{psd_confronto_rumori}(a) and Fig.~\ref{psd_confronto_rumori}(b), respectively. {The blue curves are for rectangular pulses, while the red curves are for DRAG pulses.}}
\label{150ns_confronto_rumori}
\end{figure}

This result is in stark contrast  with the comments by Ball {\sl et al.}~\cite{ball}, who instead emphasize the role of the components far from the carrier. 
{Indeed, they discuss the contribution of phase noise to fidelity degradation starting from an apparently misinterpreted consideration about the expression of the noise power spectral density}. {To convert from the phase noise to the frequency noise power spectral density, one has to multiply by $\omega^2$, which is in principle correct. However,} the frequency noise power spectral density appearing in their expression for the fidelity (Eq. (4) of Ref.~\cite{ball}) is divided by $\omega^2$, {which cancels} the effect of the previous multiplication.
On the basis of the multiplication by $\omega^2$, they state that components of the phase noise spectrum at high frequency, although lower in amplitude, have a stronger effect on fidelity than the much larger low-frequency components.
They then proceed to the calculation of the infidelity resulting from the noise from a lab-grade and from a precision local oscillator as a function of the evolution time, also including more complex sequences with dynamical error suppression. 
We believe that the results of such a calculation are most likely correct, while {just} the argument about the prevalent effect of the high-frequency components in the phase noise of the reference clock is erroneous, for the reason we just mentioned.
In order to provide further evidence for our conclusion, we have computed, with our numerical technique, the fidelity achievable for a series of 10 $\pi_x$-pulses (each 150~ns long and separated by 60~ns intervals) with a control oscillator affected by two different phase noise spectra. {The two spectra are synthesized so as to obtain one case in which the low-frequency noise is strongly prevalent, and one in which the low-frequency contribution is decreased while the high-frequency one is significantly increased.}
In particular, we consider the noise spectrum in the left panel of Fig.~\ref{psd_confronto_rumori},
where the noise level is high at low frequencies, around $-60$~dBc/Hz up to 2~MHz, and gradually decreases to approximately $-140$~dBc/Hz for frequencies above 30~MHz. We then consider the spectrum in the right panel, where the PSD is modified by lowering the low-frequency noise plateau by 20~dBc/Hz and raising the high-frequency noise plateau by 40~dBc/Hz. 
{The results are shown in Fig.~\ref{150ns_confronto_rumori}, where the blue and red curves refer to rectangular and DRAG pulses, respectively, and the top and bottom panels correspond to the noise power spectral densities of Fig.~\ref{psd_confronto_rumori}(a) and Fig.~\ref{psd_confronto_rumori}(b).} We can therefore conclude that the high-frequency components contribute quite marginally to the loss of fidelity (in particular through residual amplitude modulation, as previously discussed), both for rectangular and for DRAG pulses. Thus, in the case of broadband noise, we do not observe the reduced sensitivity found for narrow-band noise around the Rabi frequency.

{While all previously presented simulations were based on a simplified two-level energy model, we now extend our analysis to a more realistic three-level system with an anharmonicity of 300~MHz. This allows us to properly account for the leakage of population outside of the computational subspace. We applied a sequence of twelve 50~ns $\pi_x$-pulses, using both rectangular and DRAG shapes. We introduce phase noise with a Gaussian PSD, a bandwidth of 2.1~MHz, and a power level of $-80$~dBc/Hz. The analysis focuses on two specific offset frequencies: 300~MHz, matching the qubit anharmonicity, and 10~MHz, matching the Rabi frequency. 
For the 10~MHz offset we obtain a fidelity, at the end of the pulse sequence, of 0.9 and 0.8 for the DRAG and rectangular pulses, respectively. This is mainly due to the fact that the DRAG pulses suppress the leakage into the $|2\rangle$ state. For the 300~MHz offset, instead, we get about the same fidelity: 0.91 for the DRAG pulses and 0.9 for the rectangular ones, since in this case leakage is primarily caused by the noise component being exactly at the same frequency as the anharmonicity.}

\section*{Conclusions}\label{sec_three}

We investigated the effect of phase noise that affects control signals for superconducting qubits. A numerical method was developed to generate baseband phase noise values with suitable spectral characteristics, which were then added to the instantaneous phase of the carrier for the control pulses used in the simulations. The dynamics of the qubit and the time evolution of the state were simulated with Qiskit-Dynamics, which solves the time-dependent Schr\"odinger equation for the Hamiltonian describing the interaction between the qubit and the electromagnetic field.

We performed a detailed analysis of the effect of individual frequency components of phase noise on the fidelity of the final states. Starting from the analytical expression of a carrier affected by phase noise, we derived an approximate formulation that allowed us to identify the contributions related to the different types of noise modulation of the carrier. Finally, we extended the analysis by simulating the effect of a real microwave oscillator and of oscillators with a custom PSD designed to exhibit specific spectral characteristics. 

From our results, we conclude that the components that contribute most significantly to the degradation of qubit performance, and thus to the reduction in operational fidelity, are the ones with a displacement with respect to the carrier close to the Rabi frequency, and, in the case of realistic phase noise spectra, those at very low displacement, due to their large relative amplitude, which become progressively more important as the duration of the pulse sequence is increased. Instead, noise components at high displacement from the carrier give only a minor contribution, mainly limited to the residual amplitude modulation, because of their strongly suppressed power spectral density level.

\section*{Acknowledgement}
\addcontentsline{toc}{section}{Acknowledgement}

This work was partially supported by the Italian Ministry of the University and Research (MUR) in the framework of the CrossLab and the FoReLab projects (Departments of Excellence). Financial support is also acknowledged from the European Union, Next-GenerationEU, National Recovery and Resilience Plan (NRRP), Mission 4 Component 2 Investment N. 1.4, CUP N. I53C22000690001, through the National Centre for HPC, Big Data and Quantum Computing (“Spoke 10: Quantum Computing”). This work was supported also by the U.S. Department of Energy, Office of Science, National Quantum Information Science Research Centers, Superconducting Quantum Materials and Systems Center (SQMS), under Contract No. 89243024CSC000002 and by QUART\&T, a project funded by the Italian Institute of Nuclear Physics (INFN) within the Technological and Interdisciplinary Research Commission (CSN5).

\section*{Author contributions}
\addcontentsline{toc}{section}{Author contributions}

A.B. wrote the numerical codes, performed the numerical simulations, contributed to the analytical model and collected and processed the output data, G.P. provided the theoretical support for designing the method for the generation baseband phase noise sequences, P.M. contributed to the simulation codes and to processing the output data. M.M conceived and supervised the project. All authors contributed to the preparation of the manuscript and approved the final version.

\section*{Data availability}
\addcontentsline{toc}{section}{Data availability}

A set of sample codes is available as supplementary material.

\section*{Competing interests}
\addcontentsline{toc}{section}{Competing interests}

All authors declare no competing interests.

\section*{Declarations}

\subsection*{Ethics approval}
Not applicable.

\subsection*{Funding}
This work was partially supported by the Italian Ministry of University and Research (MUR) in the framework of the CrossLab and FoReLab projects (Departments of Excellence). 
Further support was provided by the U.S. Department of Energy, Office of Science, National Quantum Information Science Research Centers, Superconducting Quantum Materials and Systems Center (SQMS), under Contract No. 89243024CSC000002, and by QUART\&T, a project funded by the Italian National Institute for Nuclear Physics (INFN) within the Technological and Interdisciplinary Research Commission (CSN5). 
Agata Barsotti was supported by a PhD scholarship funded by the European Union – NextGenerationEU under the National Recovery and Resilience Plan (NRRP), within the project “PNRR - HPC, Big Data and Quantum Computing (CN1 Spoke 10: Quantum Computing)”.


\begin{thebibliography}{999}


\bibitem{Nielsen_Chuang_2010} 
M. A. Nielsen and I. L. Chuang,
{\em Quantum Computation and Quantum Information: 10th Anniversary Edition}
(Cambridge University Press, Cambridge, 2010).

\bibitem{PhysRevApplied.3.044009}
P. J. J. O'Malley, J. Kelly, R. Barends, B. Campbell, Y. Chen, Z. Chen, B. Chiaro, A. Dunsworth, A. G. Fowler, I.-C. Hoi, E. Jeffrey, A. Megrant, J. Mutus, C. Neill, C. Quintana, P. Roushan, D. Sank, A. Vainsencher, J. Wenner, T. C. White, A. N. Korotkov, A. N. Cleland, and J. M. Martinis,
Qubit Metrology of Ultralow Phase Noise Using Randomized Benchmarking,
{\em Phys. Rev. Applied} {\bf 3}, 044009 (2015).

\bibitem{Burnett2019DecoherenceBO}
J. J. Burnett, A. Bengtsson, M. Scigliuzzo, D. Niepce, M. Kudra, P. Delsing, and J. Bylander,
Decoherence benchmarking of superconducting qubits,
{\em npj Quantum Inf.} {\bf 5}, 54 (2019).

\bibitem{Carroll:2021ltj}
M. Carroll, S. Rosenblatt, P. Jurcevic, I. Lauer, and A. Kandala,
Dynamics of superconducting qubit relaxation times,
{\em npj Quantum Inf.} {\bf 8}, 132 (2022).


\bibitem{shor1995scheme}
P. W. Shor,
Scheme for reducing decoherence in quantum computer memory,
{\em Phys. Rev. A} {\bf 52}, R2493(R) (1995).

\bibitem{ryan2021realization}
C. Ryan-Anderson, J. G. Bohnet, K. Lee, D. Gresh, A. Hankin, J. P. Gaebler, D. Francois, A. Chernoguzov, D. Lucchetti, N. C. Brown, T. M. Gatterman, S. K. Halit, K. Gilmore, J. A. Gerber, B. Neyenhuis, D. Hayes, and R. P. Stutz
Realization of real-time fault-tolerant quantum error correction,
{\em Phys. Rev. X} {\bf 11}, 041058 (2021).

\bibitem{caune2024demonstratingrealtimelowlatencyquantum}
L. Caune, L. Skoric, N. S. Blunt, A. Ruban, J. McDaniel, J. A. Valery, A. D. Patterson, A. V. Gramolin, J. Majaniemi, K. M. Barnes, T. Bialas, O. Bu\v{g}dayc{\i}, O. Crawford, G. P. Geh\'er, H. Krovi, E. Matekole, C. Topal, S. Poletto, M. Bryant, K. Snyder, N. I. Gillespie, G. Jones, K. Johar, E. T. Campbell, and A. D. Hill,
Demonstrating real-time and low-latency quantum error correction with superconducting qubits,
\text{https://arxiv.org/abs/2410.05202}.


\bibitem{10.1063/1.5089550}
P. Krantz, M. Kjaergaard, F. Yan, T. P. Orlando, S. Gustavsson, and W. D. Oliver,
A quantum engineer's guide to superconducting qubits,
{\em Appl. Phys. Rev.} {\bf 6}, 021318 (2019).

\bibitem{Huang:2020sce}
H.-L. Huang, D. Wu, D. Fan, and X. Zhu,
Superconducting quantum computing: a review,
{\em Sci. China Inf. Sci.} {\bf 63}, 180501 (2020).

\bibitem{GU20171}
X. Gu, A. F. Kockum, A. Miranowicz, Y.-X. Liu, and F. Nori, 
Microwave photonics with superconducting quantum circuits,
{\em Phys. Rep.} {\bf 718-719}, 1--102 (2017).

\bibitem{doi:10.1126/sciadv.abi6690}
E. J. Zhang, S. Srinivasan, N. Sundaresan, D. F. Bogorin, Y. Martin, J. B. Hertzberg, J. Timmerwilke, E. J. Pritchett, J.-B. Yau, C. Wang, W. Landers, E. P. Lewandowski, A. Narasgond, S. Rosenblatt, G. A. Keefe, I. Lauer, M. B. Rothwell, D. T. McClure, O. E. Dial, J. S. Orcutt, M. Brink, and J. M. Chow,
High-performance superconducting quantum processors via laser annealing of transmon qubits,
{\em Sci. Adv.} {\bf 8}, eabi6690 (2022).


\bibitem{10.1063/1.525634}
G. M. Huang, T. J. Tarn, and J. W. Clark,
On the controllability of quantum‐mechanical systems,
{\em J. Math. Phys.} {\bf 24}, 2608--2618 (1983).

\bibitem{Wiseman_Milburn_2009}
H. M. Wiseman and G. J. Milburn,
{\em Quantum Measurement and Control}
(Cambridge University Press, Cambridge, 2009).

\bibitem{Dong_2010}
D. Dong and I. R. Petersen,
Quantum control theory and applications: a survey,
{\em IET Control Theory Appl.} {\bf 4}, 2651--2671 (2010).

\bibitem{aroch}
A. Aroch, S. Kallush, and R. Kosloff,
Mitigating Fast Controller Noise in Quantum Gates Using Optimal Control Theory,
{\em Quantum} {\bf 8}, 1482 (2024).


\bibitem{Kofman_2001}
A. G. Kofman and G. Kurizki,
Universal Dynamical Control of Quantum Mechanical Decay: Modulation of the Coupling to the Continuum,
{\em Phys. Rev. Lett.} {\bf 87}, 270405 (2001).

\bibitem{PhysRevLett.93.130406}
A. G. Kofman and G. Kurizki,
Unified Theory of Dynamically Suppressed Qubit Decoherence in Thermal Baths,
{\em Phys. Rev. Lett.} {\bf 93}, 130406 (2004).

\bibitem{PhysRevA.58.2733}
L. Viola and S. Lloyd,
Dynamical suppression of decoherence in two-state quantum systems,
{\em Phys. Rev. A} {\bf 58}, 2733--2744 (1998).

\bibitem{Viola_1999}
L. Viola, E. Knill, and S. Lloyd,
Dynamical Decoupling of Open Quantum Systems,
{\em Phys. Rev. Lett.} {\bf 82}, 2417--2421 (1999).

\bibitem{Uhrig_2007}
G. S. Uhrig,
Keeping a Quantum Bit Alive by Optimized $\pi$-Pulse Sequences,
{\em Phys. Rev. Lett.} {\bf 98}, 100504 (2007).

\bibitem{Khodjasteh_2009}
K. Khodjasteh and L. Viola,
Dynamically Error-Corrected Gates for Universal Quantum Computation,
{\em Phys. Rev. Lett.} {\bf 102}, 080501 (2009).

\bibitem{Biercuk_2009}
M. J. Biercuk, H. Uys, A. P. VanDevender, N. Shiga, W. M. Itano, and J. J. Bollinger,
Optimized dynamical decoupling in a model quantum memory,
{\em Nature} {\bf 458}, 996--1000 (2009).

\bibitem{filterfunction}
\L{}. Cywi\ifmmode \acute{n}\else \'{n}\fi{}ski, R. M. Lutchyn, C. P. Nave, and S. Das Sarma,
How to enhance dephasing time in superconducting qubits
{\em Phys. Rev. B} {\bf 77}, 174509 (2008).

\bibitem{Biercuk_2011}
M. J. Biercuk, A. C. Doherty, and H. Uys,
Dynamical decoupling sequence construction as a filter-design problem,
{\em J. Phys. B: At. Mol. Opt. Phys.} {\bf 44}, 154002 (2011).

\bibitem{Green_2012}
T. Green, H. Uys, and M. J. Biercuk,
High-Order Noise Filtering in Nontrivial Quantum Logic Gates,
{\em Phys. Rev. Lett.} {\bf 109}, 020501 (2012).

\bibitem{Green_2013}
T. J. Green, J. Sastrawan, H. Uys, and M. J. Biercuk,
Arbitrary quantum control of qubits in the presence of universal noise,
{\em New J. Phys.} {\bf 15}, 095004 (2013).

\bibitem{maloney}
V. Maloney, Y. Oda, G. Quiroz, B. D. Clader, and L. M. Norris, Qubit control noise spectroscopy with optimal suppression of dephasing, {\em Phys. Rev. A} {\bf 106}, 022425 (2022).

\bibitem{ball}
H. Ball, W. D. Oliver, and M. J. Biercuk,
The role of master clock stability in scalable quantum information processing,
{\em npj Quantum Information} {\bf 2}, 16033 (2016).

\bibitem{vandijk}
J. P. G. van Dijk, E. Kawakami, R. N. Schouten, M. Veldhorst, L. M. K. Vandersypen, M. Babaie, E. Charbon, and F. Sebastiano,
Impact of Classical Control Electronics on Qubit Fidelity,
{\em Phys. Rev. Appl.} {\bf 12}, 044054 (2019).

\bibitem{matsuoka}
R. Matsuoka, Y. Wachi, R. Tsuchiya, H. Mizuno, and T. Kodera,
Two-dimensional mapping method for evaluation of phase-locked loop signal in cryo-CMOS qubit control circuits,
{\em Jpn. J. Appl. Phys.} {\bf 64}, 076503 (2025).

\bibitem{jiang}
X. Jiang, J. Scott, M. Friesen, and M. Saffman,
{Sensitivity of quantum gate fidelity to laser phase and intensity noise,}
{\em Phys. Rev. Appl.} {\bf 107}, 042611 (2023) 

\bibitem{nakav}
H. Nakav, R. Finkelstein, L. Peleg, N. Akerman, and R. Ozeri,
{Effect of fast noise on the fidelity of trapped-ion quantum gates,}
{\em Phys. Rev. Appl.} {\bf 107}, 042622 (2023) 


\bibitem{qiskit}
D. Puzzuoli, J. W. Christopher, J. E. Daniel, R. Benjamin, and U. Kento,
Qiskit Dynamics: A Python package for simulating the time dynamics of quantum systems,
{\em J. Open Source Softw.} {\bf 8}, 5853 (2023).

\bibitem{qutip}
B. Li, S. Ahmed, S. Saraogi, N. Lambert, F. Nori, A. Pitchford, and N. Shammah, 
Pulse-level noisy quantum circuits with QuTip,
{\em Quantum}, {\bf 6}, 630 (2022)

\bibitem{tong}
Y. Tong and Q. Chen,
Analytical Modeling of Multiple Co-Existing Inaccuracies in RF Controlling Circuits for Superconducting Quantum Computing, {\em IEEE Trans. Comput.-Aided Design Integr. Circuits Syst.} {\bf 43}, 319--323 (2023).



\bibitem{papi2mtc}
A. Barsotti, S. Di Pascoli, and M. Macucci,
Noise Characterization and Optimization in a System for the Measurement of the Coherence Time of Superconducting Qubits,
{\em 2024 IEEE International Instrumentation and Measurement Technology Conference (I2MTC)}, 1--6 (2024).


\bibitem{10.1007/978-3-031-71518-1_12}
A. Barsotti, G. Procissi, C. Ciaramelletti, P. Marconcini, L. Guidoni, S. Paganelli, and M. Macucci,
Effect of Phase Noise in Superconducting Qubit Control,
{\em Lecture Notes in Electrical Engineering} {\bf 1263} {\em (Proceedings of SIE 2024)}, 94--101 (2025).







\bibitem{Kasdin1995}
N.~J. Kasdin,
``Discrete Simulation of Colored Noise and Stochastic Processes
and $1/f^{\alpha}$ Power Law Noise Generation,''
\textit{Proceedings of the IEEE}, vol.~83, no.~5,
pp.~802--827, 1995.

\end{thebibliography}
\end{document}